\documentclass[a4paper]{jpconf}
\usepackage{graphicx}
\usepackage{subfigure}
\begin{document}
\title{New models for PIXE simulation with Geant4}

\author{M G Pia$^1$,
	G~Weidenspointner$^{2,3}$,
        M~Augelli$^4$,
	L~Quintieri$^5$,
        P~Saracco$^1$,
        M~Sudhakar$^1$,	
        and~A~Zoglauer$^6$}

\address{$^1$ INFN Sezione di Genova, Via Dodecaneso 33, 16146 Genova, Italy}
\address{$^2$ Max-Planck-Institut f\"ur
	extraterrestrische Physik, Postfach 1603, 85740 Garching, Germany}
\address{$^3$
	MPI Halbleiterlabor, Otto-Hahn-Ring 6, 81739 M\"unchen, Germany}
\address{$^4$ CNES), 18 Av. Edouard Belin, 31401 Toulouse, France}
\address{$^5$ INFN Laboratori Nazionali di Frascati,
	Via E. Fermi 40, I-00044 Frascati, Italy}
\address{$^6$ Space Sciences Laboratory,
	University of California at Berkeley, 7 Gauss Way, Berkeley, CA 94720,
	USA}

\ead{MariaGrazia.Pia@ge.infn.it}

\begin{abstract}
Particle induced X-ray emission (PIXE) is a 
physical effect that is not yet adequately modelled in Geant4.  
The current status as in Geant4 9.2 release is reviewed and new 
developments are described. 
The capabilities of the software prototype are illustrated in 
application to the shielding of the X-ray detectors 
of the eROSITA telescope on the upcoming Spectrum-X-Gamma space mission.
\end{abstract}

\section{Introduction}

The wide use of PIXE as an experimental technique has motivated 
several dedicated software systems; nevertheless,
limited functionality for
PIXE simulation is available in general-purpose Monte Carlo codes. 
Geant4 \cite{g4nim,g4tns} addresses X-ray
emission induced both by electrons and heavy particles like protons
and $\alpha$ particles.

The simulation of PIXE involves the 
energy loss and scattering of the incident
particle, atomic shell ionization cross sections, and atomic
transition probabilities and energies.
Intrinsically, PIXE is a discrete process: X-ray emission occurs as the
result of a vacancy in shell occupancy, in
competition with Auger electron emission and Coster-Kronig transitions.
Nevertheless, this discrete process is intertwined with the ionization 
process, which determines the production of the vacancy; this process
is treated in general-purpose
Monte Carlo codes with mixed condensed and discrete transport schemes,
since excessive computational resources would be necessary to handle the
infra-red divergent cross section with a discrete scheme.

While this combined condensed and discrete particle
transport scheme is appropriate to many
simulation applications, it suffers from drawbacks with respect to the
generation of PIXE. 
Atomic relaxation occurs only 
in connection with the discrete part of the transport scheme, where
the production of a $\delta$-ray can be associated with the
creation of a vacancy;
therefore the fluorescence yield depends on the threshold 
for the secondary electron production.
Another drawback is that the cross section for the production of a $\delta$-ray
is calculated from a model for energy loss that is independent of the shell
where the ionization occurs.

This paper analyzes tools pertinent to PIXE simulation available in Geant4 at the time of the CHEP 2009 conference and describes a set of developments to extend the simulation capabilities in this domain.
A more extensive account can be found in \cite{tnspixe}.

\section{Current Geant4 tools for PIXE simulation}

Software models for the simulation of PIXE induced by protons and
$\alpha$ particles are currently available in the Geant4 low
energy electromagnetic package \cite{Chauvie2001, Chauvie2004}.

\subsection{First set of models}

The first development cycle \cite{relax_nss2004}  involved a 
software design  for interchangeable ionization cross sections models.
Based on the cross section calculations, the ionisation process creates
a vacancy, and uses Geant4 Atomic
relaxation \cite{relax} for producing fluorescence photons and
Auger electrons.
Three ionization cross section models were implemented.

The \textit{G4hShellCrossSection} class
implements cross sections based on Gryzinski's model
\cite{Gryzinski1965a,Gryzinski1965b}. 
It appears to be affected by a severe problem
of inconsistency with experimental data.

\begin{figure}[!t]
\centering
\includegraphics[width=8.5cm]{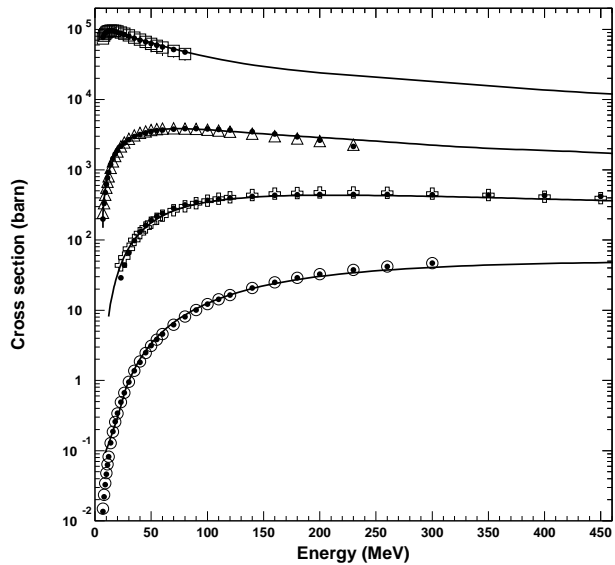}
\caption{A comparison of K shell ionization cross sections,
implemented in \textit{G4hShellCrossSectionDouble}
\cite{saliceti_tesi, mantero_phd} (black circles), with Paul and Bolik 
\cite{paul_bolik} empirical reference for $\alpha$ particles (open symbols): 
the curves are respectively for Si (squares), Cu (triangles),
Cd (crosses) and Au (white circles).
The line represents ECPSSR K shell cross sections for $\alpha$ particles
from the prototype developments of this paper.}
\label{fig_saliceti}
\end{figure}

K shell ionization cross sections based on fits to tabulated data are
implemented \cite{saliceti_tesi,mantero_phd} in 
\textit{G4hShellCrossSectionExp} and 
\textit{G4hShellCrossSectionDoupleExp}.
These models have been documented by their implementers 
\cite{saliceti_tesi,mantero_phd} as cross sections for K shell 
ionization by protons based on Paul and Sacher compilation 
\cite{paul_sacher}.
Nevertheless, they appear to describe the cross sections
for K shell ionization by $\alpha$ particles based on by Paul and Bolik
compilation \cite{paul_bolik}, as it is evident in Fig. \ref{fig_saliceti}.

\begin{figure}
\centerline{\includegraphics[width=8.5cm]{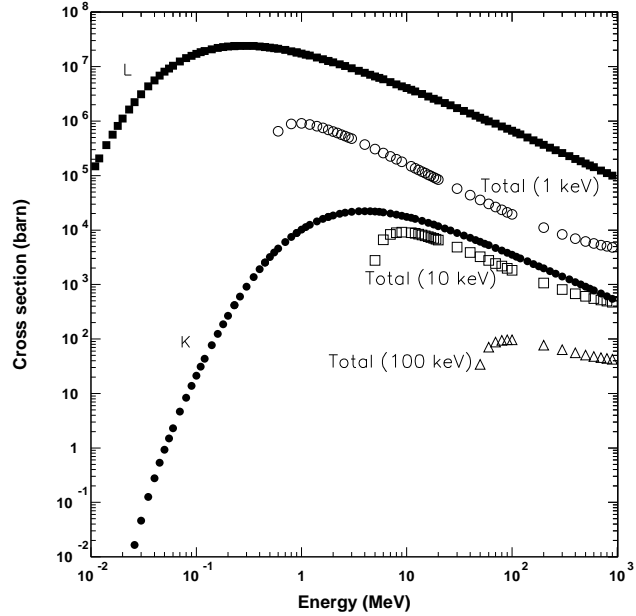}}
%
%
%
%
\caption{The filled symbols are Si ionization cross sections from an ECPSSR 
calculation: K
shell (black circles) and L shell summed over all sub-shells (black squares).
The total cross section for the discrete part of ionization as
computed in Geant4 is depicted with open symbols for different
$\delta$-ray production thresholds: 1 keV (white circles), 10 keV
(white squares) and 100 keV (white triangles).
For some $\delta$-ray production
thresholds the total cross section results smaller
than the cross section for ionising any shell.}
\label{fig_cut}
\end{figure}

The algorithm implemented to identify the shell where a vacancy is created
introduces a dependency on the user-defined electron production threshold. 
This is prone to generate inconsistent behavior, 
as it is illustrated in Fig. \ref{fig_cut}.
A detailed discussion of this issue can be found in \cite{tnspixe}.

\subsection{Models in Geant4 9.2}

A set of ionization cross section implementations \cite{haifa} for
PIXE simulation were released in Geant4 9.2.
The cross sections for K shell ionization
are based on interpolations of empirical data compilations
(Paul and Sacher \cite{paul_sacher} for protons, Paul and Bolik
\cite{paul_bolik} for $\alpha$ particles) and on the ECPSSR theory
\cite{ecpssr}; they are respectively implemented in the Geant4
\textit{G4PaulKCrossSection} and \textit{G4ecpssrCrossSection}
classes.
The cross sections for L shell ionization by protons
are based on the semi-empirical model of \cite{orlic_semiemp}.

The plots  in \cite{haifa} compare Geant4 9.2 cross section implementations 
for K
shell ionization by protons and proton
ionization experimental data, with the cross section models 
for $\alpha$ particles of the first development cycle: this
comparison is not pertinent to demonstrate the capabilities of the 
Geant4 9.2 implementations with respect to the previous ones.

The software described in \cite{haifa} and released in Geant4 9.2 
is affected by several drawbacks.

\textit{G4PaulKCrossSection}  
is slower in execution by a factor 40 to 60 approximately
than the previously existing classes implementing the same
functionality,  and two orders of magnitude slower than
the equivalent implementation described in this paper.
Other drawbacks are a memory leak and the attempt to open data files not
included in the Geant4 9.2 release.

\begin{figure}
\centering
\includegraphics[width=8.5cm]{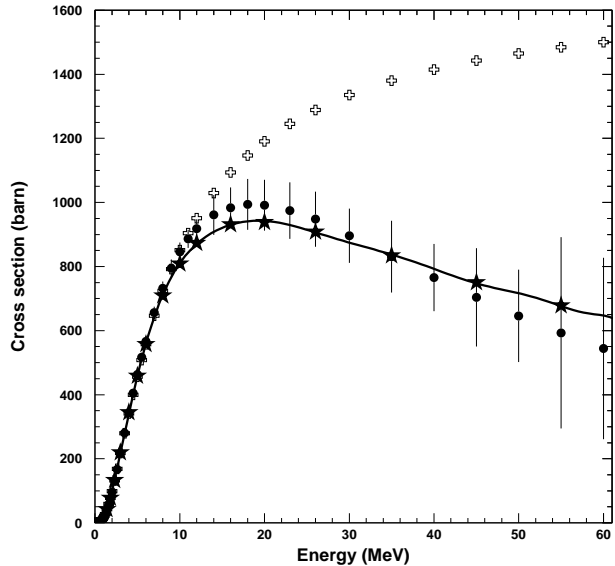}
\caption{Cross sections for copper K shell ionization by protons:
reference ECPSSR tabulations reported in \cite{paul_sacher} (stars),
ECPSSR as calculated by the \textit{G4ecpssrCrossSection} class
released in Geant4 9.2 \cite{haifa} (white crosses),
ECPSSR model as calculated by the software described in section 
\ref{sec_newpixe} based on ISICS tabulations  (solid line), and 
empirical model of Paul and Sacher \cite{paul_sacher} (black circles).
\textit{G4ecpssrCrossSection} does not appear to implement the ECPSSR 
theoretical model correctly. }

\label{fig_haifa_k29}
\end{figure}

The \textit{G4ecpssrCrossSection} class computes K shell ionization cross
section for protons and $\alpha$ particles nominally based on the ECPSSR
formulation \cite{ecpssr}.
However, it does not appear to reproduce ECPSSR cross sections correctly: for
instance, Fig. 1 to 3 in \cite{haifa} exhibit an evident divergence at higher
energies of these implementations with respect to the empirical data by Paul 
and Sacher.
Figure \ref{fig_haifa_k29} shows that this discrepancy hints to a flawed
software implementation of the ECPSSR theory  in
\textit{G4ecpssrCrossSection}, rather than a deficiency of the ECPSSR theory as
suggested by the authors of \cite{haifa}.
This class results in slower execution
(by a factor 6-8) than the ECPSSR models described in this paper.

The semi-empirical model by Orlic et al. \cite{orlic_semiemp} implemented in
\textit{G4OrlicLCrossSection} has limited capabilities:
it computes cross sections for L$_{1,2,3}$ sub-shells only for
elements with atomic number greater than 40 and 
is valid only for proton energies up to approximately 10~MeV.
Moreover, as demonstrated in
section \ref{sec_lvalidation}, it is less accurate than other available 
models.

The cross section models released in Geant4 9.2 exhibit severe software design
limitations, that prevent client code from exploiting either
dynamic or static polymorphism.
Due to their design features, these models are not practically usable
for PIXE simulation applications based on Geant4; their implementation
features - correctness, computational performance and physical
accuracy - motivate concerns about their use in experimental studies.

\section{Prototype developments for PIXE simulation with Geant4}
\label{sec_newpixe}

The new developments for PIXE simulation involved a 
re-design of the software, extended physics capabilities, the 
validation against experimental data and the application to an experimental 
use case.

\begin{figure*}
\centering
\includegraphics[width=15cm]{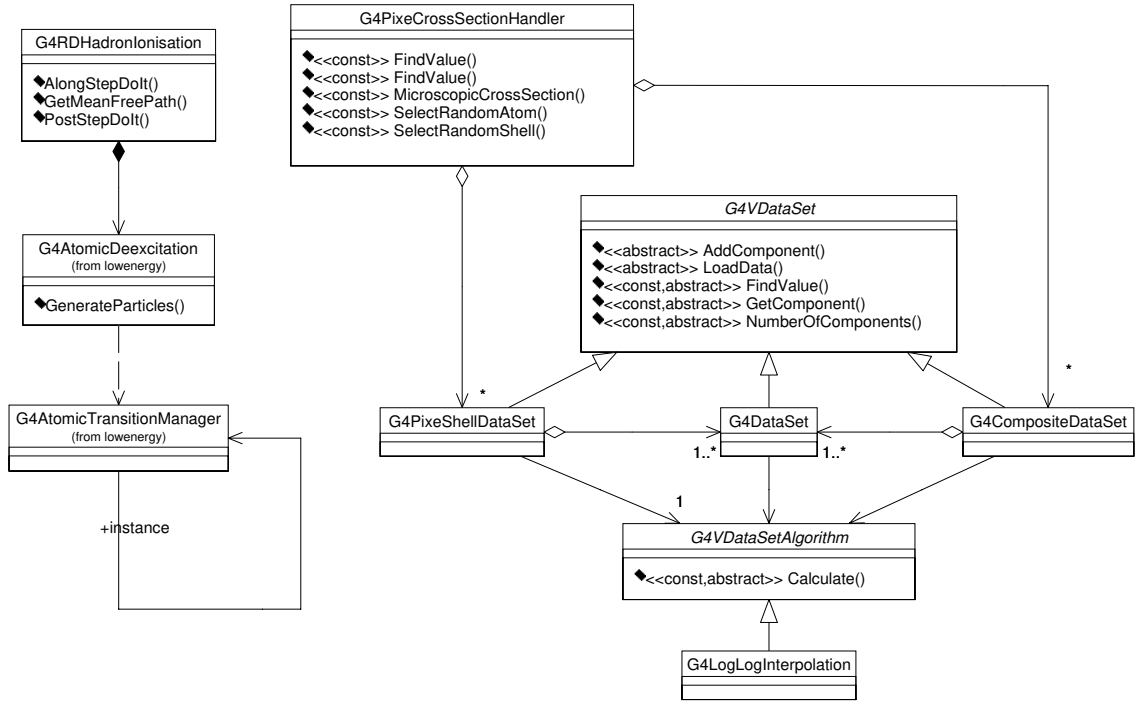}
\caption{The Unified Modelling Language class diagram
of the developments for PIXE simulation described in this paper, 
illustrating the main features of the software.}
\label{UML_fig}
\end{figure*}

The domain decomposition at the basis of PIXE simulation with Geant4
identified three main entities with associated responsibilities: the hadron
ionization process, the creation of a vacancy in the shell occupancy 
resulting from ionisation, the deexcitation of the ionised atom with 
the associated generation of X-rays.
The simulation of PIXE is the result of the collaboration of these entities.
A class diagram in the Unified Modelling Language (UML) 
illustrates the main features of the software design in Fig.~\ref{UML_fig}.

\subsection{Ionisation cross section models}
\label{sec_cross}

A wide choice of cross section models for K, L and M shell ionization
is provided in the prototype software for protons and $\alpha$ particles.
The availability of ionization cross section calculations and experimental
data for outer shells is very limited in literature.
Theoretical cross section models
include Plane Wave Born Approximation (PWBA) and variants of
the ECPSSR model: the original ECPSSR formulation \cite{ecpssr},
ECPSSR with United Atom correction (ECPSSR-UA) \cite{isics_ua}, 
ECPSSR with corrections for the Dirac-Hartree-Slater nature of the K
shell \cite{lapicki2005} (ECPSSR-HS), as well as calculations based on
recent improvements to K shell cross section specific to high energy
\cite{lapicki2008} (ECPSSR-HE).  

The cross sections 
have been tabulated and assembled in a data library; the values at a given 
energy are calculated by interpolation.
The tabulations corresponding to theoretical calculations span the
energy range between 10~keV and 10~GeV; empirical models are tabulated
consistently with their energy range of validity.
The adopted  data-driven approach optimizes performance speed and 
offers flexibility for chosing a cross section model.

ECPSSR tabulations have been produced using the ISICS software
\cite{isics,isics2006}, 2006 version and an extended version 
\cite{isics2008} including recent high energy developments.  
Tabulations of ECPSSR calculations as reported in
\cite{paul_sacher} are also provided.

Empirical cross section models for K shell
ionization include the tabulations for protons documented in
\cite{paul_sacher} and a more recent one \cite{kahoul_k}.
An empirical cross section model for K shell ionization by $\alpha$ 
particles is based on the tabulations in \cite{paul_bolik}.
Empirical models for L shell ionization by protons
have been developed by Miyagawa et al. \cite{miyagawa}, Sow et al. 
\cite{sow} and Orlic et al. \cite{orlic_semiemp}.

The ISICS software allows the calculation of cross sections for heavier
ions as well; 
therefore, the current  PIXE simulation capabilities 
can be easily extended in future development cycles.

\subsection{Generation of a vacancy}
\label{sec_vacancy}

The determination of which atomic (sub-)shell is ionised is
related to its ionisation cross section with respect to the total
cross section for ionising the target atom.
However, as previously discussed, the condensed-random-walk
scheme raises an issue as to estimating the total ionisation cross section at a
given energy of the incident particle.

A different algorithm has been adopted with respect 
to the one implemented in the first development cycle: 
the vacancy in the shell occupancy is determined based on 
the total cross section calculated by summing all the individual shell 
ionisation cross sections.
This algorithm provides a correct distribution of the produced vacancies
as long as  ionisation cross sections can be calculated for all the 
atomic shells involved in the atomic structure of the target element.
Since cross section models are currently available for K, L and M
shells only, at the present status of the software this algorithm
overestimates PIXE for elements whose atomic structure involves outer
shells, because of the implicit underestimation of the total ionization 
cross section.
This approach, however,  provides better control on the simulation results
than the algorithm implemented in the first development cycle.

The production of secondary particles by the atomic
relaxation of an ionized atom  is delegated to the
Atomic Relaxation component.

\section{Software validation}

The availability of a wide variety of cross section models for the
first time in the same computational environment allowed a detailed
comparative assessment of their features against experimental data.
Due to the limited page allocation in the conference proceedings and 
copyright constraints with the publisher of \cite{tnspixe}, only 
a brief summary is reported here; further details can be found in \cite{tnspixe}, 
including the full list of elements subjects 
to test against experimental data and a large collection of plots 
demonstrating the comparison between simulation and experimental data.

The comparison of cross sections as a function of energy was performed
for each element by means of the $\chi^2$ test.
Contingency tables were built on the basis of the outcome of the
$\chi^2$ test to determine the equivalent, or different behavior of
model categories.
The input to contingency tables derived from the results of the
$\chi^2$ test: they were classified respectively as ``pass'' or
``fail'' according to whether the corresponding p-value was consistent
with a 95\% confidence level.
The contingency tables were analyzed with Fisher exact test
\cite{fisher}. 

\subsection{K shell ionization cross sections}
 
The reference experimental data were extracted from
\cite{paul_sacher}.

\begin{table*}
\begin{center}
\caption{Percentage of test case with compatibility at 
confidence level CL between simulation models and experimental data of K
shell ionization by protons}
\label{tab_kpass}
\begin{tabular}{c|c|c|c|c|c|c}
\hline
CL &ECPSSR &ECPSSR-HE	&ECPSSR-HS	&ECPSSR-U  &Paul-Sacher	&Kahoul\\
\hline
\multicolumn{7}{c}{All measurements} \\
\hline
95\%    	&67    &74    &77    &68    &71    &46  \\
99\%		&85    &83    &83    &85    &80    &57  \\
\hline
\multicolumn{7}{c}{Excluding high energy data} \\
\hline
95\%    	&69    &75    &86    &69    &70    &48  \\
99\%		&83    &81    &91    &83    &80    &56  \\
\hline
\end{tabular}
\end{center}
\end{table*}

The fraction of test cases for which the $\chi^2$ test fails to reject the 
null hypothesis at the 95\% and 99\% confidence level are listed in 
Table \ref{tab_kpass}:
all the cross section models implemented in the simulation exhibit
equivalent behaviour regards the compatibility with the 
experimental data, with the exception of the Kahoul 
et al. model.
The contingency table comparing the Kahoul et al.
and ECPSSR-HS models  
confirms that the two models show a statistically significant difference
regards their accuracy (p-value of 0.001).

The contingency tables associated with the other models show that they 
are statistically equivalent regards their accuracy.
However, when only the lower energy range (below 5-7~MeV, depending on
the atomic number) is considered, a statistically significant
difference at the 95\% confidence level (p-value of 0.034)
is observed between the ECPSSR
model and the ECPSSR-HS one; the latter is more accurate with
respect to experimental data. 

From this analysis one can conclude that the implemented 
K shell ionization cross
section models exhibit a satisfactory
accuracy with respect to experimental measurements.

\subsection{L shell ionization cross sections}
\label{sec_lvalidation}

The cross sections for L sub-shell ionization cross sections were compared to
the experimental data collected in two complementary compilations \cite{sokhi},
\cite{orlic_exp}.
The same method was applied as described for the validation of K shell cross
sections.

The ECPSSR model appears to provide a satisfactory representation of
L shell ionisation cross sections with respect to experimental data,
especially with its United Atom variant.

The ECPSSR-UA exhibits the best overall accuracy among the various models;
the Orlic et al. model exhibits the worst accuracy with respect to 
experimental data.  
This semi-empirical
model is the only option implemented in Geant4 9.2 for the calculation of L
shell ionization cross sections.

The accuracy of the various cross section models was studied by means of
contingency tables to evaluate their differences quantitatively.
The categorical analysis was performed between the ECPSSR model with United Atom
correction, i.e. the model showing the best accuracy according to the results of
the $\chi^2$ test, and the other cross section models.
The contingency tables were built based on the results of the $\chi^2$ test at
the 95\% confidence level, summing the ``pass'' and ``fail'' outcome over the
three sub-shells.

The Orlic et al. semi-empirical model is found to be significantly less accurate
than the ECPSSR-UA model: the hypothesis of equivalence of their accuracy with
respect to experimental data is rejected at 99\% confidence level.
The p-values concerning the comparison of the Miyagawa et al. empirical model
are close to the critical region for 95\% confidence, and slightly different for
the three tests performed on the related contingency table.
The Sow et al. empirical model and the ECPSSR model in its original formulation
appears statistically equivalent in accuracy to the ECPSSR model with United
Atom correction.

As a result of this analysis, the ECPSSR model with United Atom approximation
can be recommended for usage in Geant4-based simulation applications as the most
accurate option for L shell ionization cross sections.
The ECPSSR model in its original formulation can be considered a satisfactory
alternative; the Sow et al. empirical model has satisfactory accuracy, but
limited applicability regards the target elements and proton energies it can
handle.

\subsection{Cross section models for high energy PIXE}
\label{sec_hepixe}

PIXE as a technique for elemental analysis is usually performed with proton
beams of a few MeV.
In the recent years, higher energy proton beams of a few tens MeV have been 
effectively exploited too.
High energy protons are a source of PIXE in the space radiation 
environment.

The interest in high energy PIXE has motivated recent theoretical investigations
\cite{lapicki2008} concerning cross section calculations at higher energies.
Despite the emerging interest of high energy PIXE, only a limited set of
experimental data is available above the energy range of conventional PIXE
techniques.

The accuracy of the implemented K shell cross section models was evaluated
against two sets of measurements at higher energy
\cite{denker,pineda}, respectively at 66 and 68 MeV.
The experimental measurement with uranium was not included in the comparison,
since it appears affected by some experimental systematics.

The $\chi^2$ test was performed first separately on either experimental data set
to evaluate the possible presence of any systematics in the two test cases, then
on the combined data set.
The p-values from the $\chi^2$ test against these experimental data are listed
in Table \ref{tab_hepixe}.

Over the limited data sample considered in this test, the ECPSSR model 
with the correction in \cite{lapicki2008}this model does not appear
to provide better accuracy than the original ECPSSR formulation; nevertheless
more high energy experimental data would be required to reach a firm 
conclusion.
Also, this analysis should be verified over tabulation deriving 
from a published version of the ISICS code, when it becomes available.

\begin{table}
\begin{center}
\caption{P-values from the $\chi^2$ test concerning high energy experimental data}
\label{tab_hepixe}
\begin{tabular}{l|c|c|c|c}
\hline
Experimental		&ECPSSR &ECPSSR		&ECPSSR		&ECPSSR \\
data			&	&High Energy   	&Hartree-Slater	&United Atom\\
\hline
\cite{denker}, 68 MeV	&0.612	&0.069		&0.054		&0.612 \\
\cite{pineda}, 66 MeV	&0.235	&0.060		&$<0.001$ 	&0.235 \\
Combined		&0.351	&0.020		&$<0.001$ 	&0.351 \\
\hline
\end{tabular}
\end{center}
\end{table}

\section{Application of the PIXE prototype software}
\label{sec_application}

The prototype components for PIXE simulation described in the previous sections
were applied to a study of the passive, graded Z shielding of the X-ray 
detectors of the eROSITA telescope  \cite{erosita} on the upcoming Russian 
Spectrum-X-Gamma space mission. 

\begin{figure}[!t]
\centering
\subfigure[Cu Shield]
{\includegraphics
[width=8.0cm]{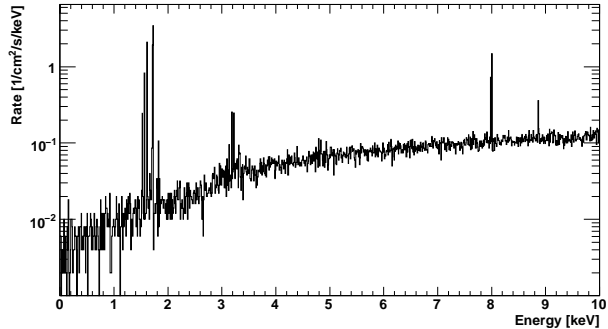}
\label{comp_Cu-shield}}
\vspace*{1ex}
\subfigure[Cu-Al Shield]{\includegraphics
[width=8.0cm]{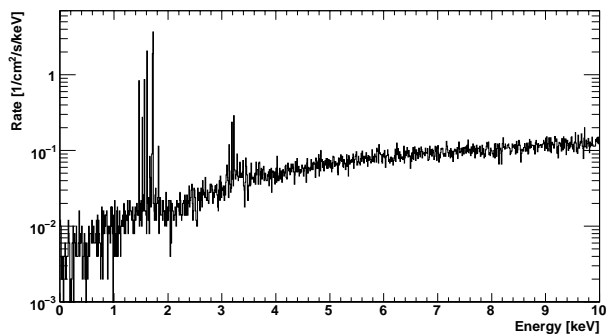}
\label{comp_Cu-Al-shield}}
\vspace*{1ex}
\subfigure[Cu-Al-B$_4$C Shield]
{\includegraphics
[width=8.0cm]{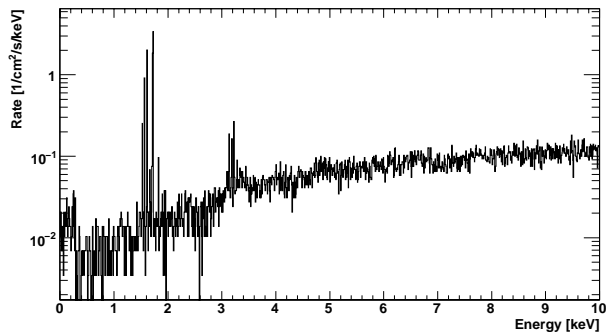}
\label{comp_Cu-Al-B4C-shield}}
\caption{A comparison of the fluorescence background due to ionization
by cosmic-ray protons in an L2 orbit for three different graded Z
shield designs for the eROSITA X-ray detectors.}
\label{comp_shield}
\end{figure}

The background spectra due to cosmic-ray protons were simulated for 
the three different eROSITA graded Z shield designs. 
A comparison of the results is depicted in Fig.~\ref{comp_shield}.

This application demonstrates that the developed software is
capable of supporting concrete experimental studies.
Nevertheless, the concerns outlined in the previous sections 
should be kept in mind: while the present PIXE simulation component
can provide valuable information in terms of relative fluorescence yields
from inner shells, the intrinsic limitations of the mixed transport scheme
in which ionization is modelled and the lack of cross section
calculations for outer shells prevent an analysis 
of the simulation outcome in absolute terms.

\section{Conclusion and outlook}
\label{summary}

This paper presents a brief overview of the status, open
issues and recent developments of PIXE simulation with Geant4;
a more extensive report of the underlying concepts, 
developments and results is available in \cite{tnspixe}.

The analysis of the related models currently available in Geant4
showed their limitations under various aspects: physics functionality,
correctness of implementation, accuracy, consistent behavior, software
design, performance and usability.

The new developments represent a significant 
step forward regards PIXE simulation with Geant4.
They extend the capabilities of the toolkit
by enabling the generation of PIXE associated with K, L and M shells 
for protons and $\alpha$ particles; for this purpose 
a variety of cross section models are provided.
The adopted data-driven strategy and the software design 
improve the computational performance over
previous Geant4 models.
The validity of the implemented models has been quantitatively
estimated with respect to experimental data.
The results provide objective guidance
for the optimal selection of simulation models in user applications.

An extensive ionisation cross section data library has been created as a
by-product of the development process: it can be of interest to the
experimental community for a variety of applications, not necessarily
limited to PIXE simulation with Geant4.

Some issues identified in the course of the development process are
still open: they concern the consistency of PIXE simulation in a mixed
condensed-discrete particle transport scheme.
A possible approach to address them, while preserving the current
continuous-discrete design scheme of hadron ionisation in Geant4,
would involve the calculation of ionization cross sections for outer
shells than M, 
This calculation is feasible, exploiting known theoretical methods, 
yet it would require a significant investment of resources.
In parallel, a project \cite{nano5} is in progress to address 
design issues concerning co-working condensed and
discrete transport methods in a general purpose simulation system.

Despite the known limitations related to mixed transport schemes, the
software developments described in this paper provide sufficient
functionality for realistic experimental investigations.
The plans concerning the public availability of these new developments are discussed in \cite{tnspixe}.

\section*{Acknowledgment}

The authors express their gratitude to A.~Zucchiatti for valuable
discussions on PIXE experimental techniques, to S.~Cipolla for providing 
a prototype version of ISICS 2008, and to
S. Bertolucci and U. Bratzler for helpful comments and advice.

The authors are grateful to the RSICC staff at ORNL, in
particular B. L. Kirk and J. B. Manneschmidt, for the support to
assemble a ionization cross section data library for public
distribution resulting from the developments described in this paper.

\end{document}